\documentclass[12pt]{spieman} 
\usepackage{amsmath,amsfonts,amssymb}
\usepackage{graphicx}
\usepackage{setspace}
\usepackage{tocloft}
\usepackage{cite,multirow,tabularx}
\usepackage{ifpdf,multicol}
\usepackage{algorithmic,balance}
\usepackage{booktabs}
\usepackage{longtable}
\usepackage{tabu}
\newcolumntype{+}{>{\global\let\currentrowstyle\relax}}
\newcolumntype{^}{>{\currentrowstyle}}

\newcolumntype{L}[1]{>{\raggedright\arraybackslash}p{#1}}
\title{Indirect Microwave Holography with Resolution Enhancement in Metallic Imaging}
\author[a]{Vineeta Kumari}
\author[a]{Aijaz Ahmed}
\author[b*]{Gyanendra Sheoran}
\author[c]{Tirupathiraju Kanumuri}
\author[d]{Chandra Shakher}
\affil[a]{Department of Electronics \& Communication Engineering, National Institute of Technology Delhi, Delhi -110040, India}
\affil[b]{Department of Applied Sciences (Physics), National Institute of Technology Delhi, Delhi -110040, India}
\affil[c]{Department of Electrical and Electronics Engineering, National Institute of Technology Delhi, Delhi -110040}
\affil[d]{SeNSE
Centre for Sensors, INstrumentation and Cyber-physical Systems Engineering,Indian Institute of Technology Delhi, Delhi-110016}

\begin{document} 

\maketitle

\begin{abstract}
The development of compact indirect microwave holographic set-up by the implementation of low cost, specifically designed directive antennae as transmitter and receiver is proposed. Microwave holograms are recorded by 2D scanning over a plane using motorized translation stage. The recorded interference pattern i.e. holograms are then processed numerically to reconstruct the amplitude and phase information employing the angular spectrum diffraction method. The quality of the reconstructed amplitude image is further enhanced through the deep neural network, in order to combat with the low resolution of reconstructed images. The qualitative experimental results exploit the possibility of developing the miniaturized, and low cost indirect microwave holographic system for near field applications.
\end{abstract}


\section{Introduction}
Microwave holography consists of the acquisition of scattered wave-field from an object and provides the amplitude and phase information of the same. 
It is a non-destructive and non-invasive technique which is based on interference while recording the hologram and diffraction principle for reconstructing the hologram. Microwave holography is categorized as - direct and indirect. In direct holography, the complex scattered field is measured over a selected aperture in the near field using Vector Network Analyzer (VNA) to image the objects. This technique is very expensive and computationally complex as it is based on vector calculations \cite{r0,r1}. On the other hand, indirect holography is a two-step process that employs the methods of recording and reconstruction of an object. The hologram is acquired in the form of interference pattern produced by the interference of the synthesized reference wave \cite{r2} and the object wave-front scattered from its surface. The whole information (i.e.amplitude and phase) of an object can be retrieved from a single hologram. In accordance of the phase shifting digital holography \cite{r3}, the phase shift of reference wave in microwave holography is introduced by using a phase shifter of application specific frequency band. 

Earlier, the indirect holographic technique was used for the determination of antenna radiation characteristics \cite{r4}. Thereafter, its usage in imaging of concealed metal objects has also been well documented in the literature \cite{r5,r6}. The technique also finds its apt use in imaging the objects in security and biomedical applications \cite{r7,r8,r9}.

 The literature for indirect holography has used open ended wave guide probe antennae for the recording of the hologram \cite{r5,r6,r7,r8,r9}. Generally, a probe antenna is bulky and has a fixed dimension for a specified frequency band; the gain of a probe antenna is fixed for a certain design and application. Thus, such types of antenna have limitations to be used in different (multiple/various) imaging applications, e.g. it cannot be well utilized in the field of medical imaging, where the required gain of the antenna varies from negative to certain positive values. The size of the whole set-up could not be reduced due to the large size of probe antenna.  
These snags in the transmitting and receiving antennas motivated the authors to utilize a small size, low cost and lightweight patch antenna in the indirect holographic system.

Previously, we have proposed the use of microstrip patch antenna for receiving the scattered wave-fields in the indirect holographic set-up with a horn antenna as transmitter \cite{r10}. Although, the results were in good agreement with the original scanned objects but it was limited due to its low directivity and narrow band range of operations. For developing an imaging system of better accuracy, the antennae need to be highly directive and of wideband range.

In this paper, the holographic arrangement using specifically designed UWB Vivaldi antennae with a maximum gain at ${45^o}$ as transmitter and receiver has been demonstrated experimentally. It has been depicted that there is considerable reduction in the size of the system. The  amplitude and phase are reconstructed effectively with the use of small size Vivaldi antennae. The vivaldi antennae have benefits over the traditionally used probe antennae in terms of their size scalability for use at multiple frequencies, cost effectiveness and broadband characteristics. Further, the resolution of the reconstructed amplitude has also been improved using a well trained deep neural network i.e. VDSR network \cite{r11}. To the best of authors' knowledge, it is the maiden attempt to explore Vivaldi antenna for use in indirect microwave holography.

\section{Materials and Methods}
The whole experimental set-up of indirect microwave holography comprises of - (i) Design and fabrication of antennae and (ii) Recording and reconstruction of holograms.
\subsection{Antenna design}
\begin{figure}[htb!]
\centering
\includegraphics[width=0.8\linewidth]{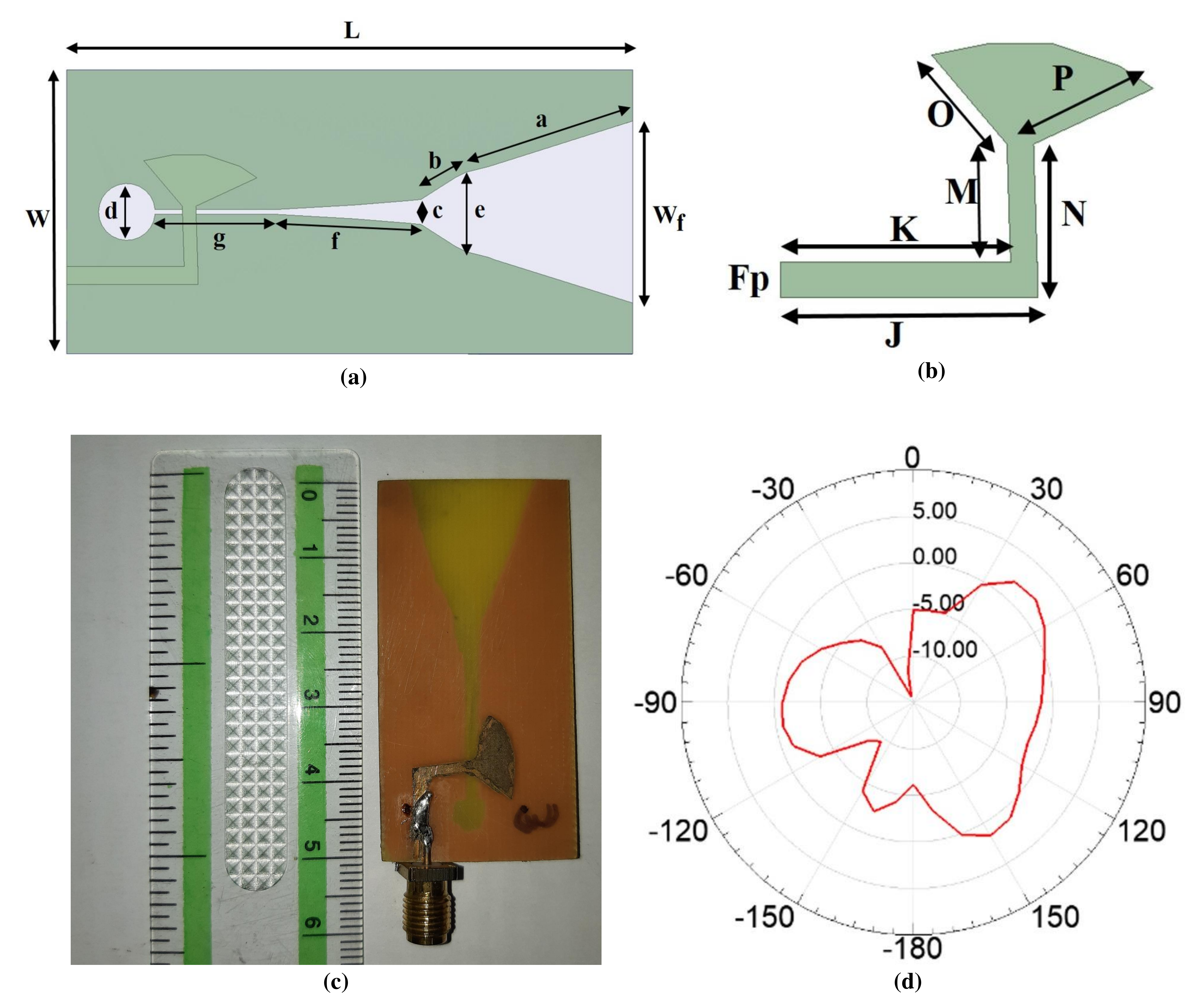}
\vspace{-0.3cm}
\caption{Schematic of the proposed antenna (a) conductive layer (b) capacitive strip feed (c) fabricated antenna (d) Directive gain of the antenna (dB)}\label{f1}
\vspace{-0.3cm}
\end{figure}
Vivaldi antenna is a kind of tapered slot UWB antenna which has wide bandwidth, high directivity and is able to produce symmetrical radiation pattern \cite{r212}. 
The proposed Vivaldi antenna is specifically designed and fabricated to work in the frequency range of 3.8-4.5 GHz and 5.2 - 17 GHz, with the peak gain of 3.53 dB at ${45^o}$. Therefore, it eliminates the design and fabrication of multiple antennae for  near field holographic imaging applications. Here, this aforesaid ${45^o}$ field directionality is achieved by introducing the non-identical flanges to the standard structure of Vivaldi antenna. A slant is provided on the one side of the flange corresponding to the other flange to achieve the additional field length along the direction of propagation. This brings in the bending of the radiation field in a particular direction. The bending of radiation is required in the proposed set-up for keeping the isolation in between the radiation fields of the transmission and receiving antennas.
The antenna structure is simulated with High-Frequency Structure Simulator (HFSS) and designed on a dielectric substrate FR-4 with a dimensions of  50mm $\times$ 25mm $\times$ 0.8 mm. 

\begin{figure}[htb!]
\centering
\includegraphics[width=0.8\linewidth]{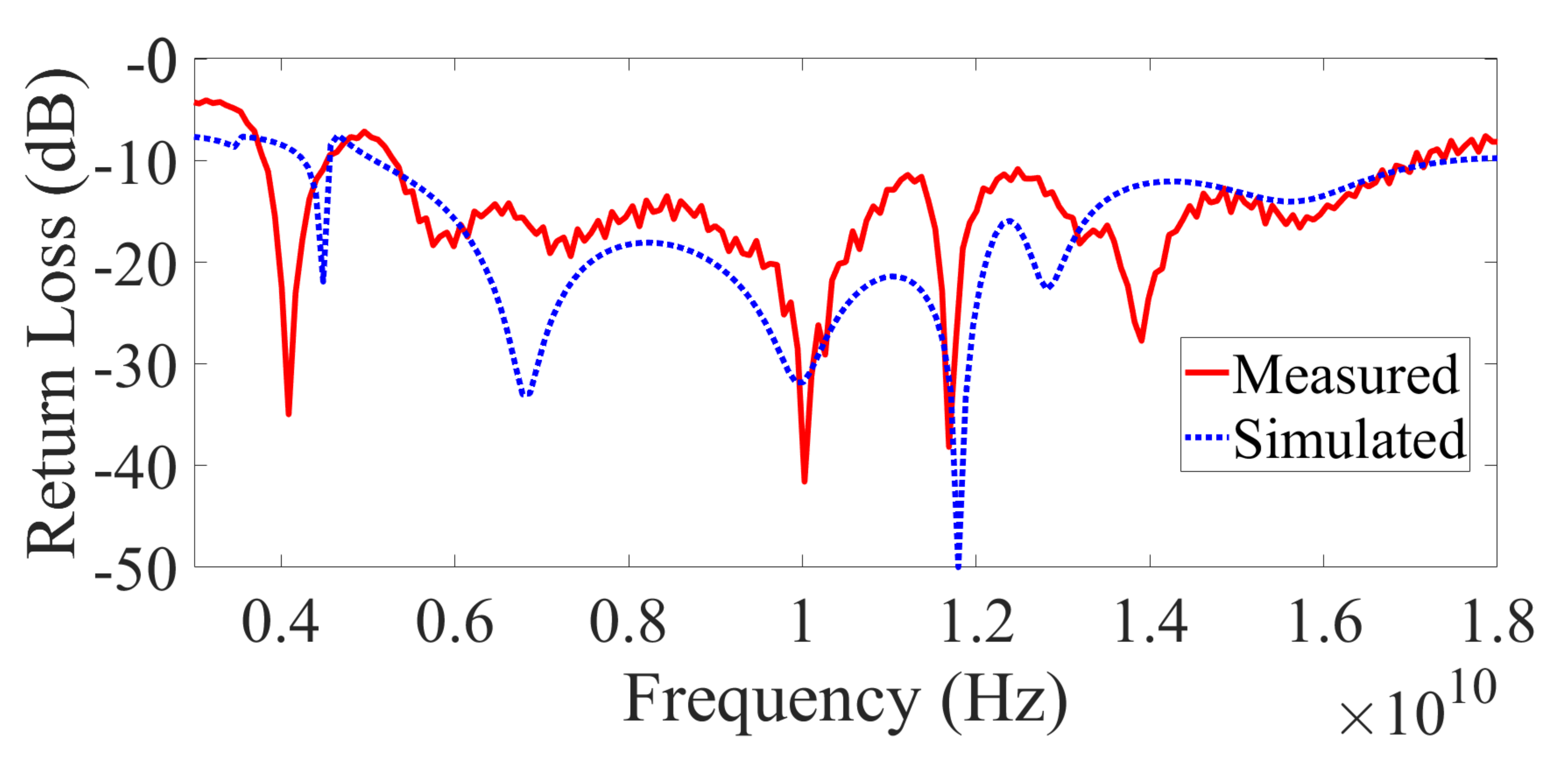}
\vspace{-0.3cm}
\caption{Comparison of the simulated and the measured return loss graph of the designed antenna }\label{f2}
\vspace{-0.3cm}
\end{figure}

Figs. \ref{f1}(a) and (b) show the schematic of the designed antenna with detailed marking of designed parameters. The designing parameters of antenna are given in Table 1. Apart from that, the closed and open ends of the conductive layer are given as 0.4 mm and 16 mm respectively and `Fp' is the 50 $\Omega$ feed line of the antenna. 
 
Fig. \ref{f1}(c) depicts the in-house fabricated antenna, here, a ruler has also been shown to define the size of the fabricated antenna. Also, the gain of the antenna is shown in Fig. \ref{f1}(d), which shows the parametric evaluation of the designed antenna to demonstrate the effect of the change in the upper flange in the direction of radiation pattern. The directivity of the radiation pattern is not solely dependent on the change of upper flange's length but it also depends on the irregularity introduced in the upper flange.



Fig. \ref{f2} compares the simulated and the experimental return loss of the antenna, which describes the simulated frequency band and experimental frequency band achieved at desired frequency range. Two identical antennae acting as a transmitter and receiver are designed and fabricated for conducting the experiments.  

\begin{figure}[htb!]
\centering
\includegraphics[width=0.8\linewidth]{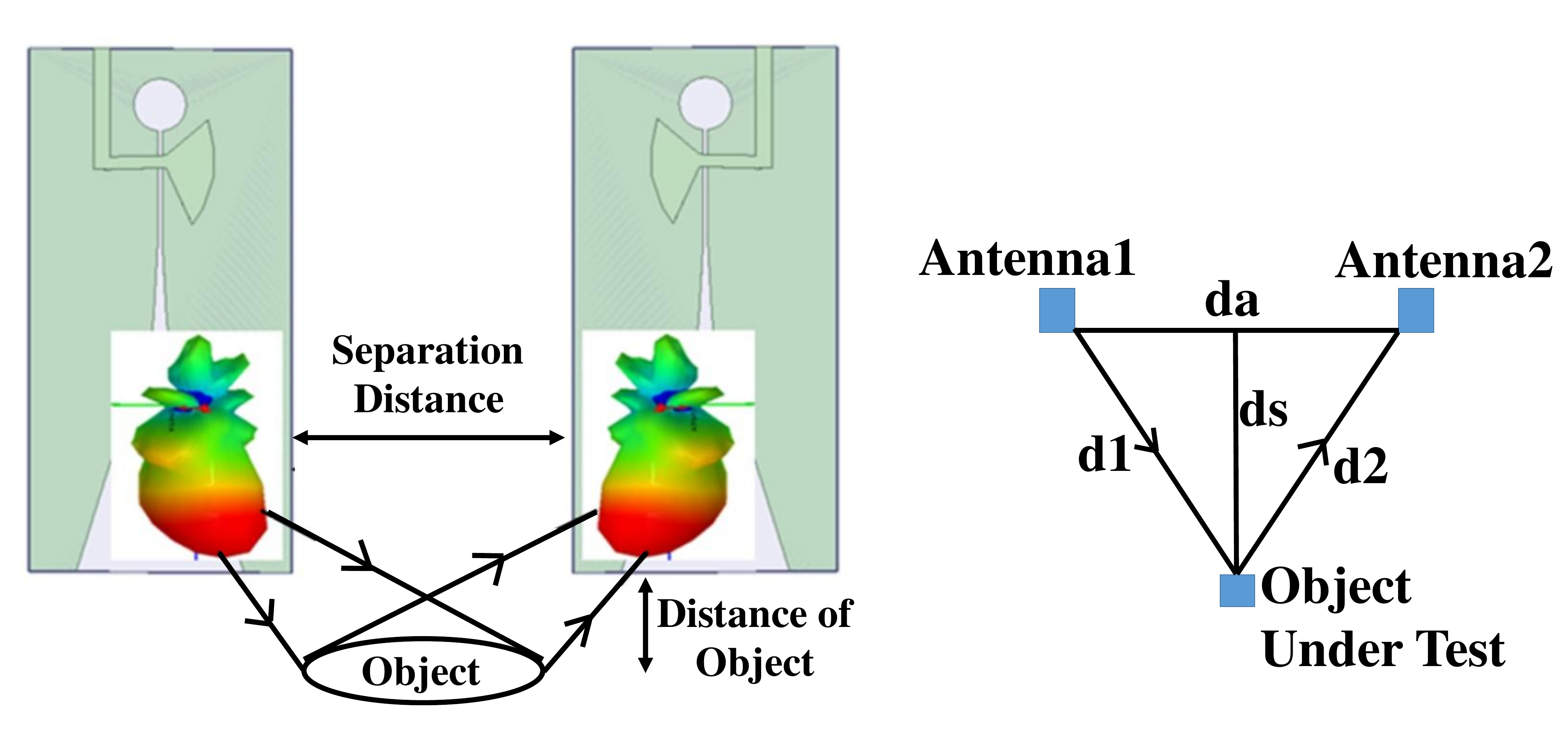}
\vspace{-0.3cm}
\caption{Conceptual arrangement of the antenna in the measurement setup.}\label{f3}
\vspace{-0.3cm}
\end{figure}

\begin{figure}[htb!]
\centering
\includegraphics[width=0.6\linewidth]{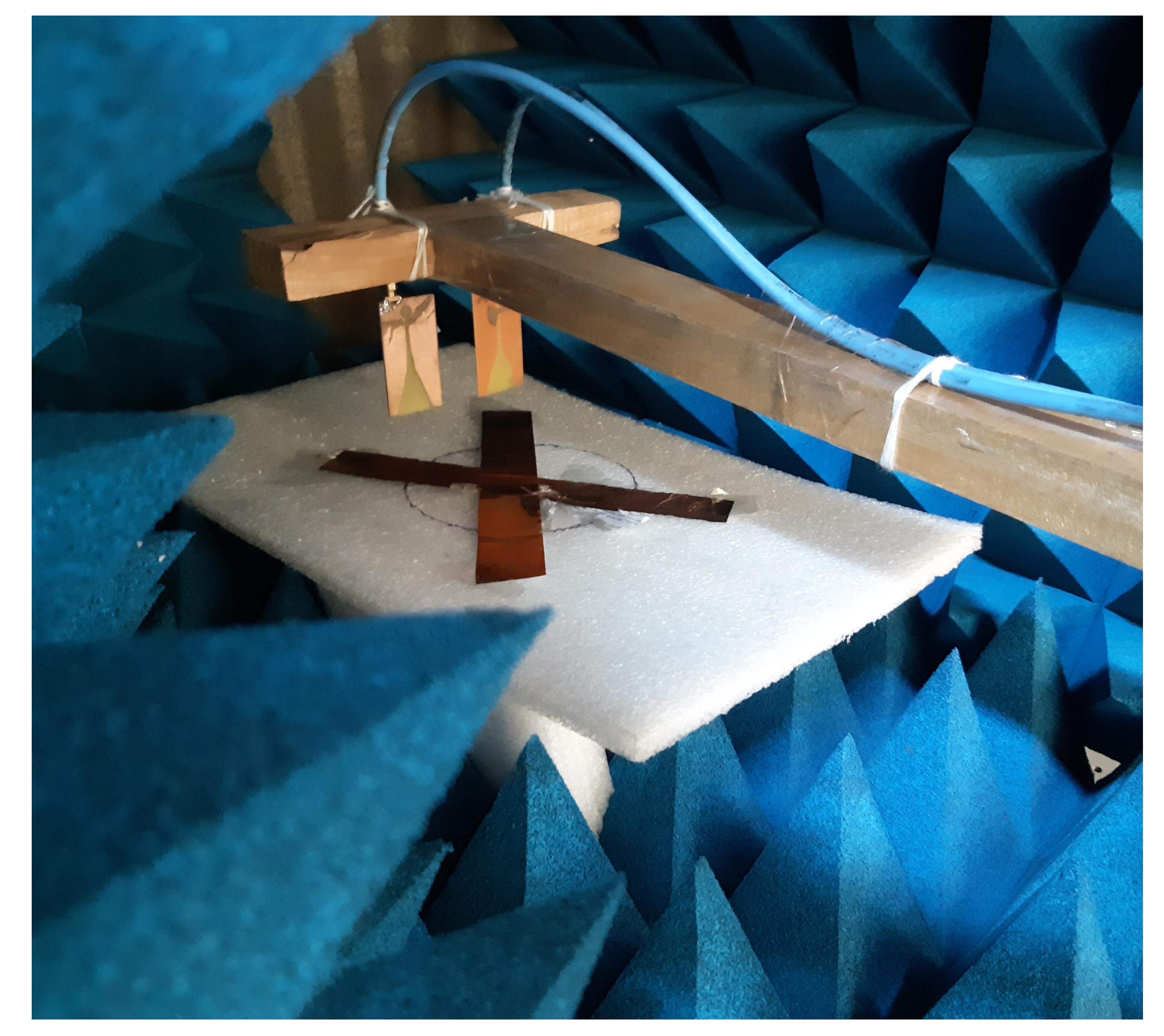}
\vspace{-0.3cm}
\caption{Setup arrangement for imaging of copper sheets with shape `X'.}\label{f4}
\vspace{-0.3cm}
\end{figure}

The conceptual arrangement of antennae in the experimental setup and its simplified form is shown in Fig. \ref{f3}. The separation between the object and the antennae is
calculated using right-angle property of triangle. The distance `ds' between the antenna and object can vary in the near-field region of the antenna. In order to maintain a specific recording distance of the object, the value of the separation distance `da' is calculated by keeping in mind that `d1' should be equal to `d2' to have a common field of view. The experimental setup arrangement for the antennae with the calculated proper separation distance from the object is shown in Fig. \ref{f4}. 
\subsection{Mathematical Formulation of Microwave Holography} 
The ability to reconstruct the interference pattern utilizing the scalar power measurement techniques is the fundamental of indirect microwave holography. By combining the complex field of the scattering wave from the object, with a synthesized reference wave \cite{r2}, of known phase and amplitude, holographic information can be derived. 
Here, the object wave is combined with a known reference wave with constant amplitude and linearly increasing phase shift. The resulting interference pattern i.e. hologram (H(x, y)) is recorded using elementary and economical power meter utilizing hybrid Tee junction. The resultant hologram or interference pattern is given in Eq.\ref{e1}:
\begin{equation}
{\rm{H(x,y)  = }}{\left| {{\rm{O(x, y) }} \pm {\rm{R(x, y) }}} \right|^2} \label{e1}
\end{equation}
Where O(x, y) defines the scattered field detected by the antenna from the object, and R(x, y) denotes the reference wave. The plus and minus signs are corresponding to the resultant intensities recorded from the sum and difference port of the Hybrid Tee respectively. This is employed in order to reduce the overlapping in between the diffraction orders in the Fourier domain.

\begin{figure}[htb!]
\centering
\includegraphics[width=0.5\linewidth]{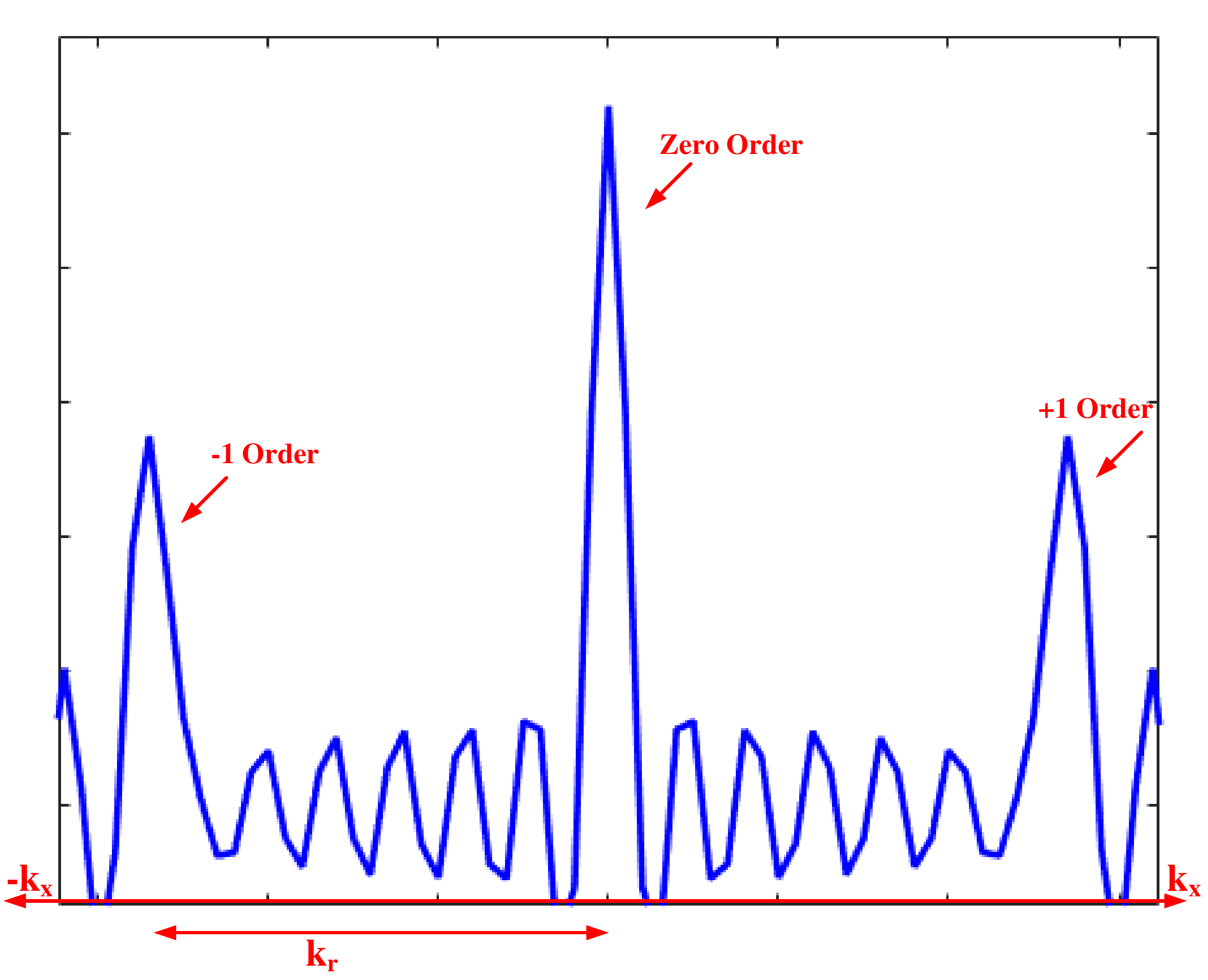}
\vspace{-0.3cm}
\caption{Frequency spectrum representation of the interferogram}\label{f5}
\vspace{-0.3cm}
\end{figure}

Taking the Fourier transform of Eq. \ref{e1}, we obtain the spectrum for interference pattern or hologram, 
\begin{equation}
\begin{array}{l}
{\rm{F\{ H(x, y)\}   =  F\{ }}{\left| {{\rm{O(x, y)}}} \right|^2}{\rm{\}  + F\{ }}{\left| {{\rm{R(x, y)}}} \right|^2}{\rm{\} }}...\\
\,\,\,\,\,\,\,\,\,\,\,\,\,\,\,\,\,{\rm{ ...+ F\{ O(x,y)}} . {{\rm{R}}^*}{\rm{(x,y)\}  + F\{ }}{{\rm{O}}^*}{\rm{(x,y)}} . R{\rm{(x,y)\} }}
\end{array} \label{e2}
\end{equation}
The hologram spectrum consists of three diffraction orders: DC term or zero order, +1 order and -1 order. Fig. \ref{f5} is representing the 1D line profile of the frequency spectrum of the recorded hologram. Here, `${{\rm{k}}_r}$' defines the offset vector occurs due to the phase sift in reference wave, where as, `${{\rm{k}}_x}$' is the function of wave propagation  in x-direction.

Here, the zero order is in the center of the frequency domain spectrum and the +1, -1 orders are shifted on the axis according to the linear phase shift of the reference wave, which is given in Eq.\ref{e3},
\begin{equation}
{\rm{R(x,y) = }}{{\rm{E}}_{\rm{0}}}{{\rm{e}}^{{\rm{ - j}}{{\rm{k}}_{\rm{r}}}{\rm{(x}}\,{\rm{ + y)}}}}\label{e3}
\end{equation}
In Eq.\ref{e3}, ${{\rm{k}}_r}$ represents the offset wave vector, which depends on the sample spacing during scanning in the x and y axis i.e. $\Delta{\rm{x}}$ and $\Delta{\rm{y}}$. This offset wave vector is used to synthesize reference wave by introducing the phase shift of $\Delta\phi = 2\pi/3$ rad between the sample spacing of $\Delta x =\lambda/6$, which produces an electronically synthesized reference wave with a offset of ${{\rm{k}}_r} \geq 2k$. The benefits of the synthetically generated reference wave is that the linear phase gradient generation enables the measurements in the near field of the antenna.

A suitable and precise sampling separation must be provided to reconstruct the object image (both amplitude and phase). 
The image reconstruction is done by a diffraction technique i.e. Angular Spectrum Method (ASM). The recorded and reconstructed pixel size of hologram remains same in ASM.
In this method, the recorded complex signal's plane wave spectrum in x-y plane is defined by Fourier spectrum as shown in Eq. \ref{e2}. The signal is back propagated to the object plane and inverse Fourier transform is performed to reconstruct the object using Eq.\ref{e4}.
\begin{equation}
\begin{array}{l}
{\rm{e(x,y)  =  FT}}_{{\rm{2D}}}^{{\rm{ - 1}}}\left[ {{\rm{F}}{{\rm{T}}_{{\rm{2D}}}}\left[ {{\rm{I}}{\rm{(x,y)}}} \right]{\rm{exp( - j}}{{\rm{z}}_{\rm{0}}}\sqrt {{\rm{4}}{{\rm{k}}^{\rm{2}}}{\rm{ - }}\,{\rm{k}}_{\rm{x}}^{\rm{2}}{\rm{ - }}\,{\rm{k}}_{\rm{y}}^{\rm{2}}} )} \right]\\\\
where,\,\,{k_x} = \frac{{2\pi }}{{N\Delta x}},\,\,{k_y} = \frac{{2\pi }}{{N\Delta y}}
\end{array}\label{e4}
\end{equation}

Where, $z_0$ is representing the distance in between the object plane and the recording plane (where
the holograms are recorded) and in between the recording and reconstruction plane, the complex
amplitude is reconstructed at $z_0=0$ i.e at object plane.

The phase of the complex amplitude is calculated by:
\begin{equation}
\phi (x,y) = {\tan ^{ - 1}}\left( {\frac{{im\,{\rm{(e(x,y))}}}}{{re\,{\rm{(e(x,y))}}}}} \right)\label{e5}
\end{equation}
The phase obtained from Eq. \ref{e5} is wrapped in the range of (-$\pi$, +$\pi$), the reconstructed phase images are shown in the results section.

\subsection{Experimental Setup}
\begin{figure}[h]
\centering
\includegraphics[width=0.6\linewidth]{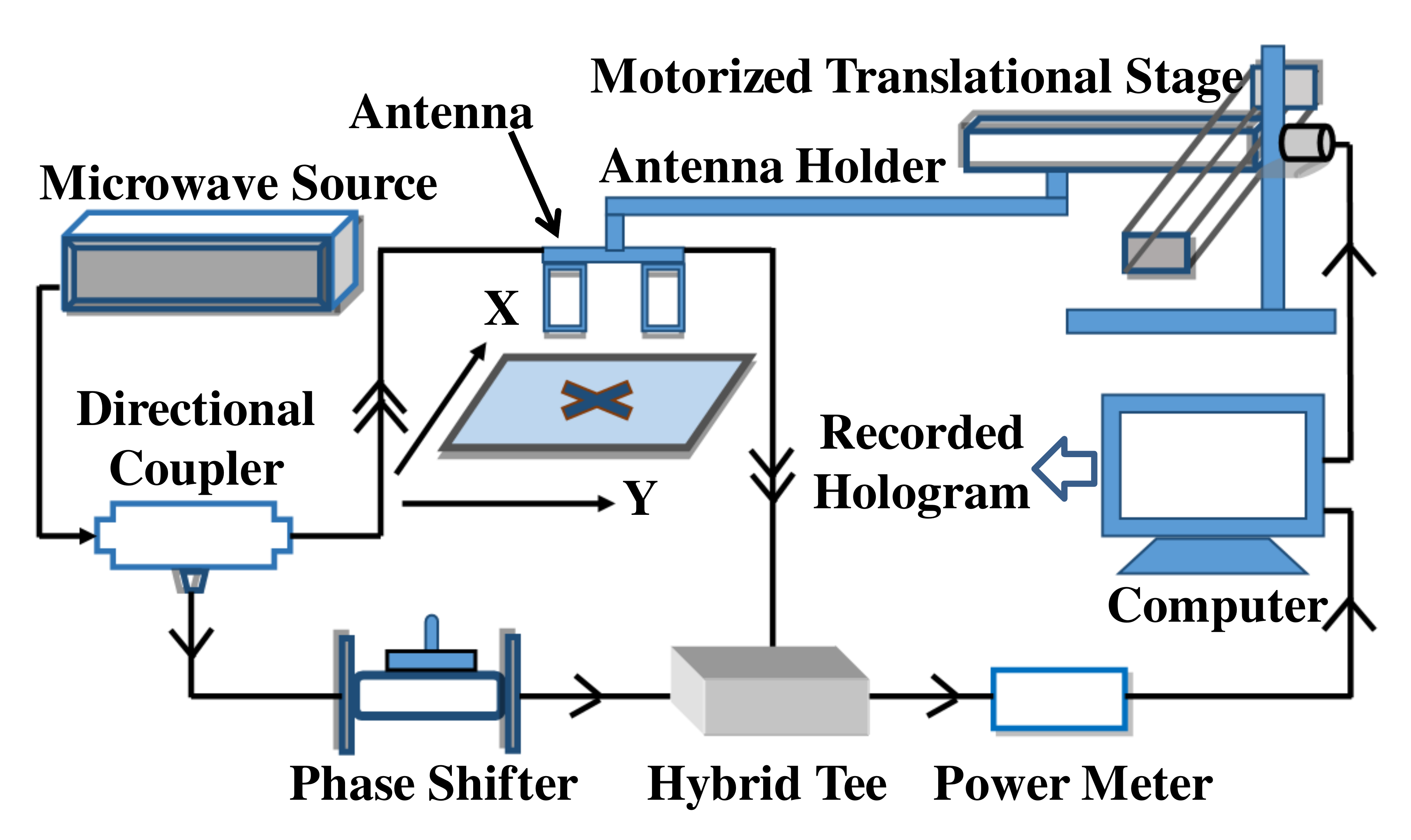}
\vspace{-0.3cm}
\caption{Schematic of the experimental setup of Near Field Indirect Microwave Holography.}\label{f6}
\vspace{0.5cm}
\end{figure} 
The schematic of the experimental microwave holographic set-up is shown in Fig.\ref{f6}. The microwave source (Rohde and Schwarz- Model No. SMB100A) is used to be operated at 9.1 GHz frequency and a power of 0 dBm while recording the hologram. 
\begin{figure}[htb!]
\centering
\includegraphics[width=0.6\linewidth]{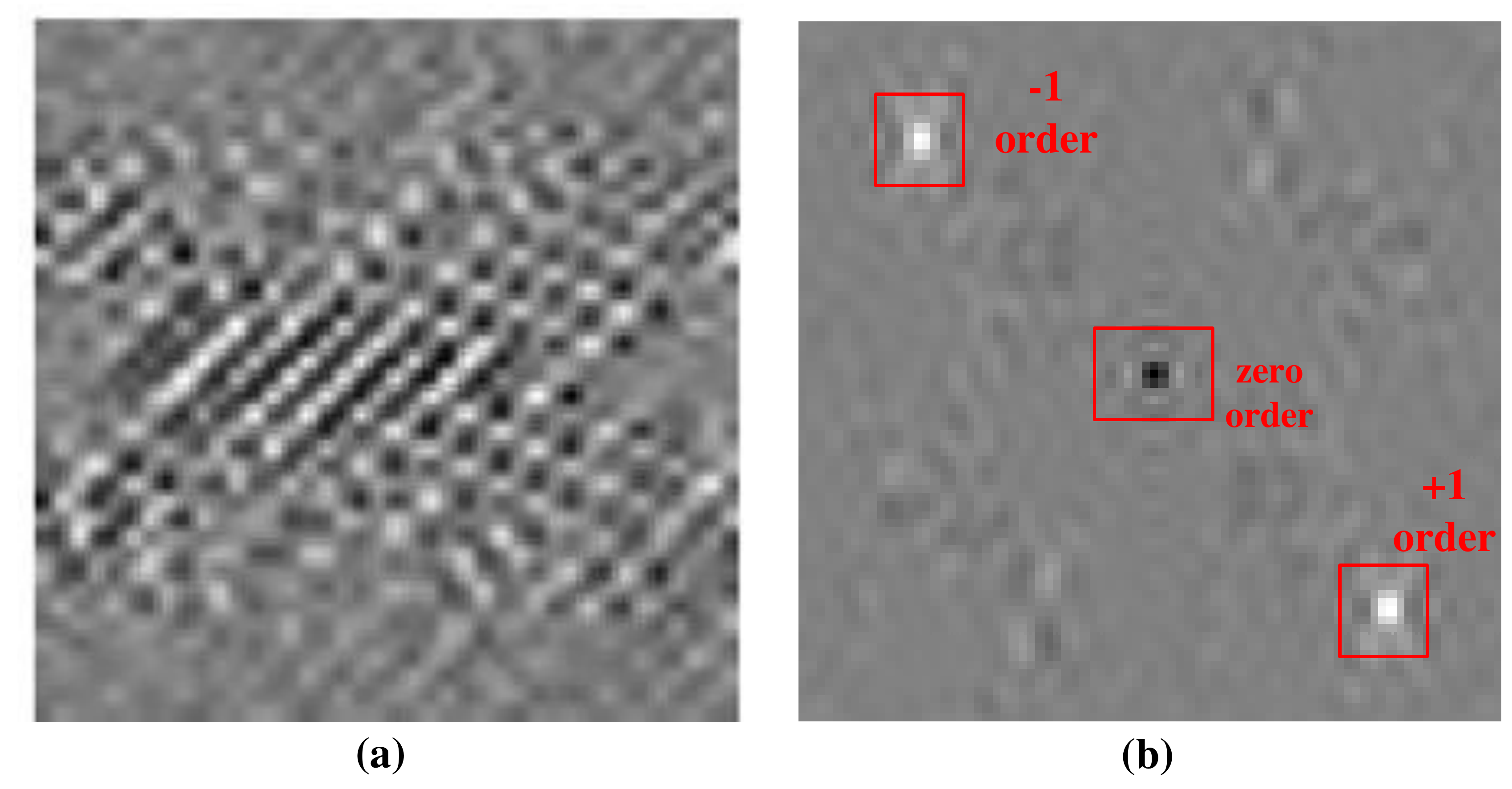}
\vspace{-0.3cm}
\caption{Isolation graph of both the antennas }\label{f71}
\vspace{-0.3cm}
\end{figure}
A microwave signal is propagated from the source to the directional coupler with coupling factor -10 db, through probe, which is used to tape off the wave into two parts. One part behaves as an object wave, which is made to fall on the object under investigation through the transmitting Vivaldi antenna. The identical transmitting and receiving antennas should be isolated to the field of each other. For isolation, the S21 and S12 parameters are to be measured and it should be below -15dB. The measured isolation for the antennas is below the desired value as shown in Fig. \ref{f71}.

The other part is forwarded from the directional coupler to the attenuator, which is used to generalize the whole experimental set-up for both low and high gain antennas based on various application. The attenuator is utilized to adjust the level of the reference and scattered signals according to the gain of the antenna and to increase the dynamic range of the hologram.
The attenuated signal is further used to generate a synthesized reference wave by providing a 3-step linear phase shift in `x' and `y' direction to separate the twin images and dc term. The receiving second antenna is used to collect the scattering wavefield from the object surface at the receiver side. This antenna is connected to a linear XY motorized translation stage to perform a 2D scan across the image plane. The sample spacing of x and y (i.e. $\Delta x$ and $\Delta y$) is taken as 5mm. The corresponding receiving power at each scanning step of the antenna is made to interfere with the reference wave and measured. The power meter, which is connected to a personal computer controlling the movement of the antennas across the scanning aperture and imports the power meter data, reads the received power consisting of the combination of the scattered signal with the tapped coherent reference signal.Holograms have been recorded with a linear phase increment of $2\pi/3$ rad for each sample spacing.

The whole scanning aperture is shielded using microwave absorbers to minimize the scattering effects of outer objects while performing the experiments. A reference hologram was recorded without placing any object to subtract the background noise from the resultant hologram.
The post processing is done by reconstructing the amplitude and phase of the recorded hologram numerically.

\subsection{Resolution Enhancement}
\begin{figure}[h!]
\centering
\includegraphics[width=0.45\linewidth]{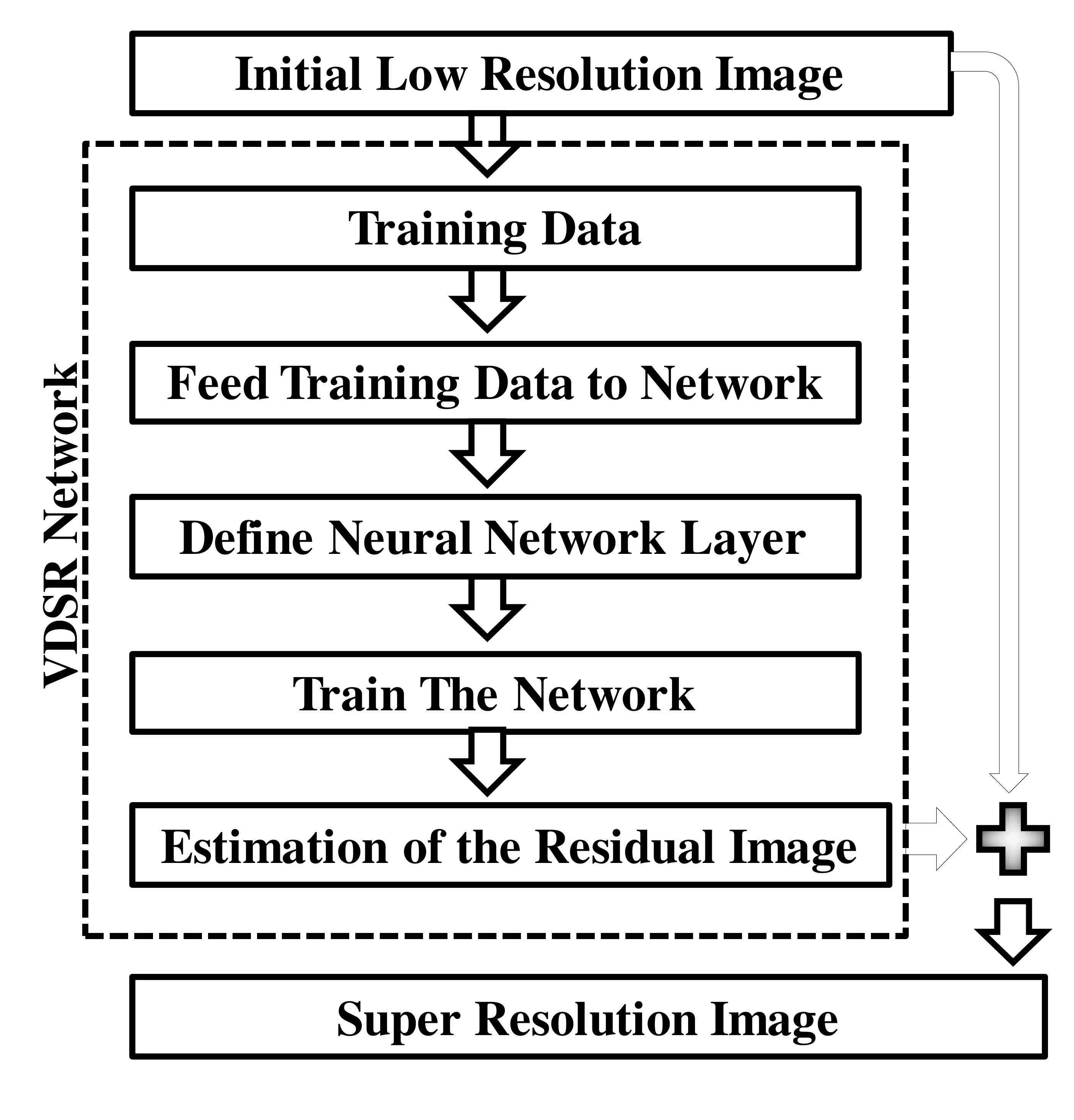}
\vspace{-0.3cm}
\caption{Process Flow of VDSR Network}\label{f81}
\vspace{-0.3cm}
\end{figure}

Since, the hologram recorded is with 40$\times$40 pixels so it is having low resolution, because the lower spatial frequencies are recorded due to the less number of samples, which are limited because of longer wavelength of the microwaves. Hence, the quality of the reconstructed amplitude images is improved from lower resolution to higher resolution by applying a well-trained VDSR network \cite{r11}, which has used a training data set of 291 images along with a network of depth 20. The loss function of the network has been given as $\frac{1}{2}{\left\| {r - f(x)} \right\|^2}$
where `$r$' is the residual image and `$f(x)$' is the network prediction \cite{r11}. 
Since, this network is having very high learning rates and better accuracy as compared to other existing networks, therefore it has been implemented for the improvement of reconstructed amplitude images. This network uses a residual image learning strategy for enhancing the resolution. The residual image depicts the variation of the low and high-resolution images and it contains the high frequency details. The network approximates the residual image and improves the image resolution by appending the same into the low-resolution images. The mechanism for improving the quality of the reconstructed amplitude images is shown in flow chart in Fig. \ref{f81}.

Subsequently, the signal to noise ratio (SNR) has also been improved. To evaluate the deep neural network technique, SNR and Structural Similarity Index (SSI) are calculated as\cite{r12}-
\begin{equation}
{\rm{Speckle}}\,{\rm{Index}} = \,\sqrt {\frac{{{\rm{local}}({\mathop{\rm var}} (\overline x ))}}{{{\rm{local}}(E(\overline {x)} )}}}\label{e6}
\end{equation}
\begin{equation}
{\rm{SNR = }}\frac{{\rm{1}}}{{{\rm{Speckle}}\,{\rm{Index}}}}\label{e7}
\end{equation}
and,
\begin{equation}
{\rm{SSI(x,y) =}} \,\frac{{(2{\mu _x}{\mu _y} + {c_1})(2{\sigma _{xy}} + {c_2})}}{{({\mu _x}^2 + {\mu _y}^2 + {c_1})({\sigma _x}^2 + {\sigma _y}^2 + {c_2})}}\label{e8}
\end{equation}

Where, $\mu_x$ and $\mu_y$ are means of images x and y respectively, $\sigma_x$ and $\sigma_y$ are variances for the respective images, $\sigma_{xy}$ shows the co-variance and ${c_1}$ and ${c_2}$ are the variables used to stabilize the division. Structural similarity index is calculated for the resolution enhanced images to define that there are no structural variations after applying the neural network. The calculated values of aforesaid evaluation measures for resolution enhanced object images  are shown in Table II.
Since, there is a very sharp transition in phase values wrapped in ranging from - $\pi$ to + $\pi$, therefore approximation method may distort the phase values and hence the wrapped phase can't be unwrapped without error. Hence, to avoid any phase error, the resolution of the reconstructed wrapped phase images is not improved with neural network method. 

\section{Results and Discussion}
The performance of the proposed experimental set-up is demonstrated on 4 types of metallic objects having different shapes, sizes and scattering properties. The object is placed at 25mm from the scanning aperture. Fig. \ref{f4} shows the experimental arrangement of placing an object in the vicinity of the transmitted and receiving antenna. The preliminary tests have been carried out on two thin copper sheets of size 185mm $\times$ 25mm, each sheet is arranged in `X' shape as object. 

\begin{figure}[htb!]
\centering
\includegraphics[width=0.75\linewidth]{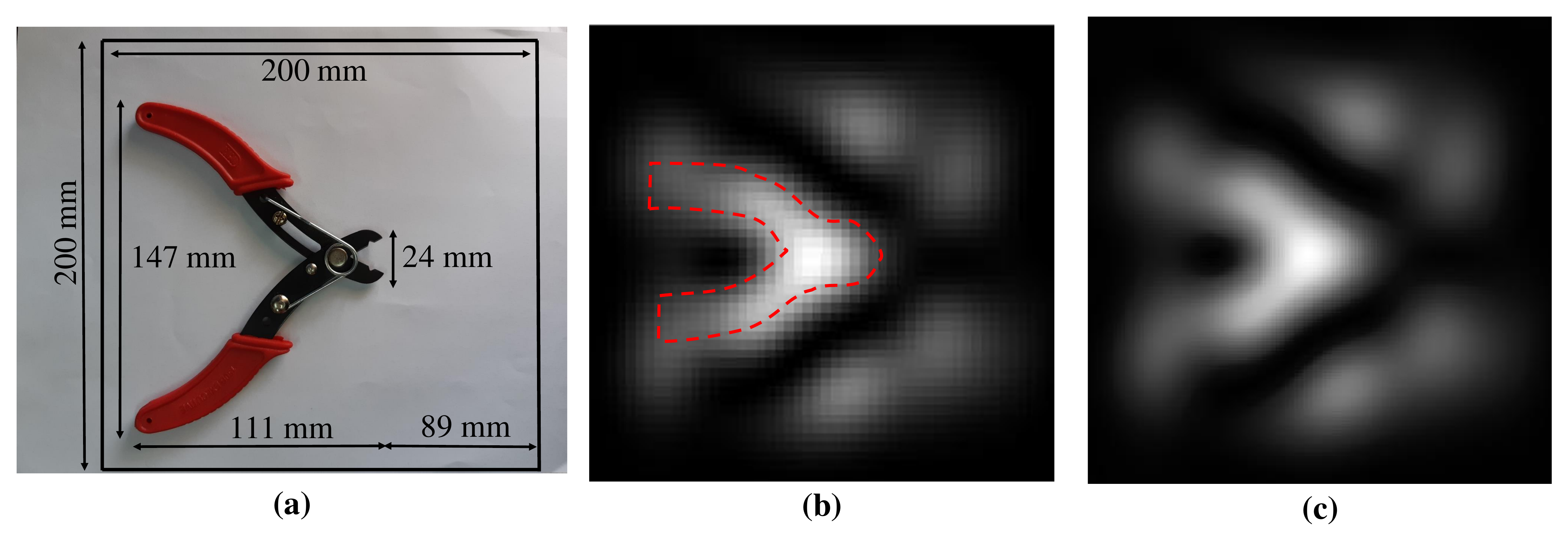}
\vspace{-0.3cm}
\caption{(a) Recorded hologram of the object (b) Frequency spectrum.}\label{f7}
\vspace{-0.3cm}
\end{figure}

Fig. \ref{f7}(a) shows the recorded 3-step phase shifted hologram consisting of 40 $\times$ 40 pixels, after taking its Fourier transform, all the frequency orders are separated as +1 order, -1 order and zero order representing the real image, virtual image and DC term respectively as shown in Fig. \ref{f7}(b).

\begin{figure}[htb!]
\centering
\includegraphics[width=\linewidth]{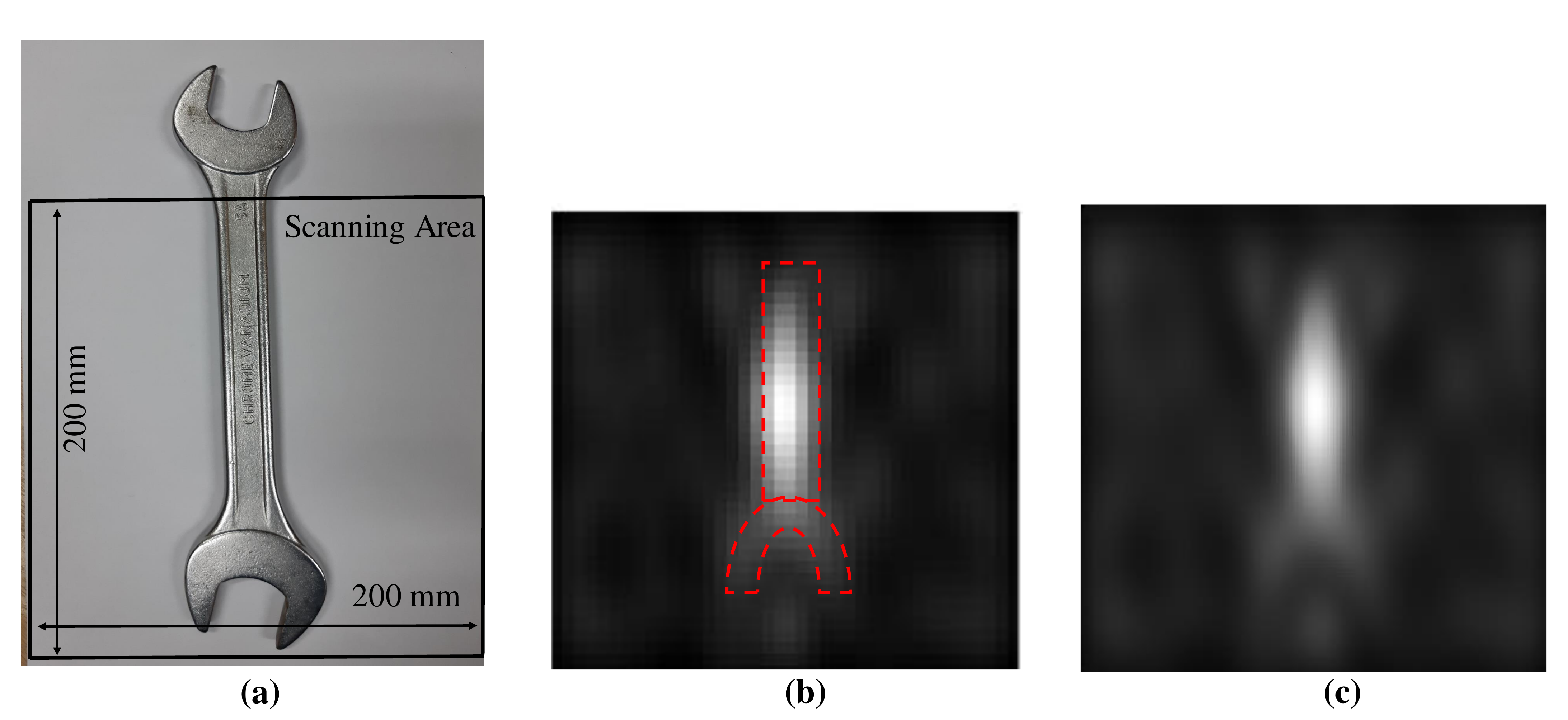}
\vspace{-0.3cm}
\caption{(a) Reconstructed amplitude image of `X' copper strip (b) Resolution enhanced image (c) Reconstructed wrapped phase Image.}\label{f8}
\vspace{-0.3cm}
\end{figure}

\begin{figure}[htb!]
\centering
\includegraphics[width=\linewidth]{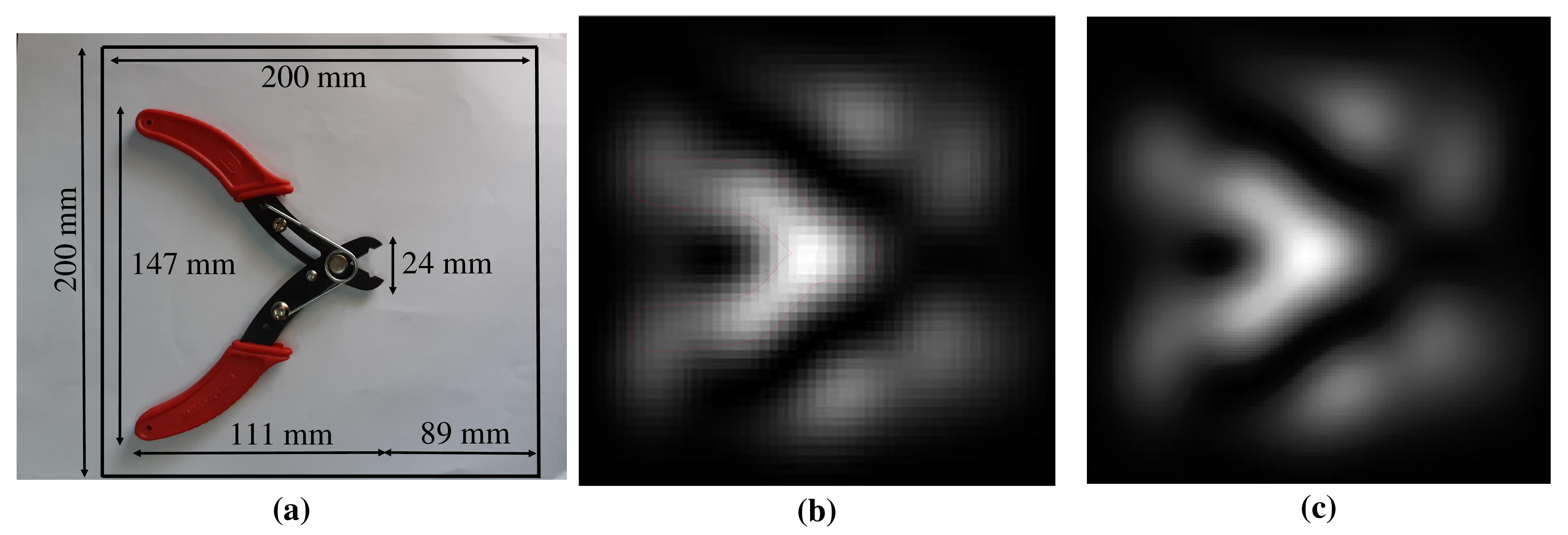}
\vspace{-0.3cm}
\caption{(a) Object image (b) Reconstructed amplitude image of wire scrapper as object (c) Resolution enhanced image.}\label{f9}
\vspace{-0.3cm}
\end{figure}

\begin{figure}[htb!]
\centering
\includegraphics[width=\linewidth]{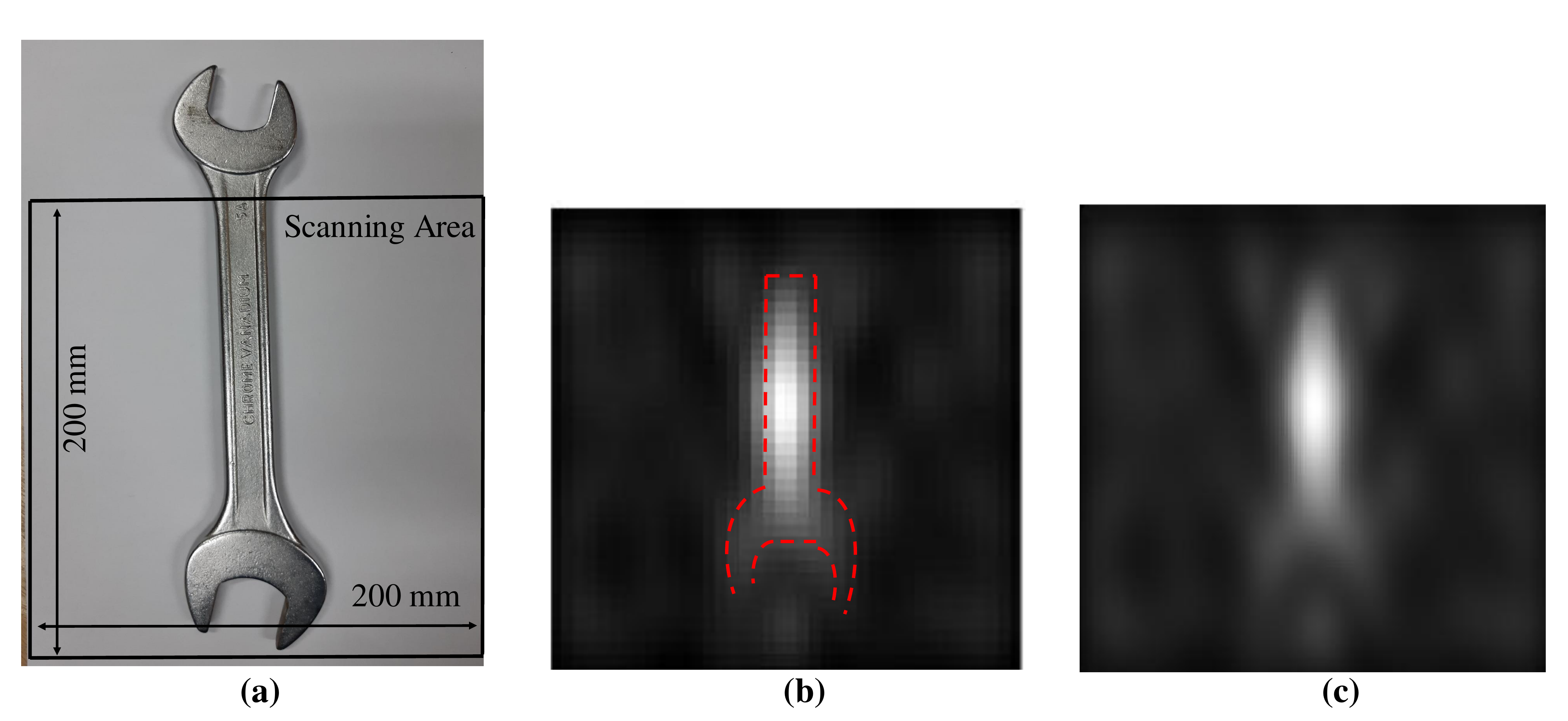}
\vspace{-0.3cm}
\caption{(a) Object image (b) Reconstructed amplitude image of spanner as object (c) Resolution enhanced image.}\label{f10}
\vspace{-0.3cm}
\end{figure}

\begin{figure}[htb!]
\centering
\includegraphics[width=0.8\linewidth]{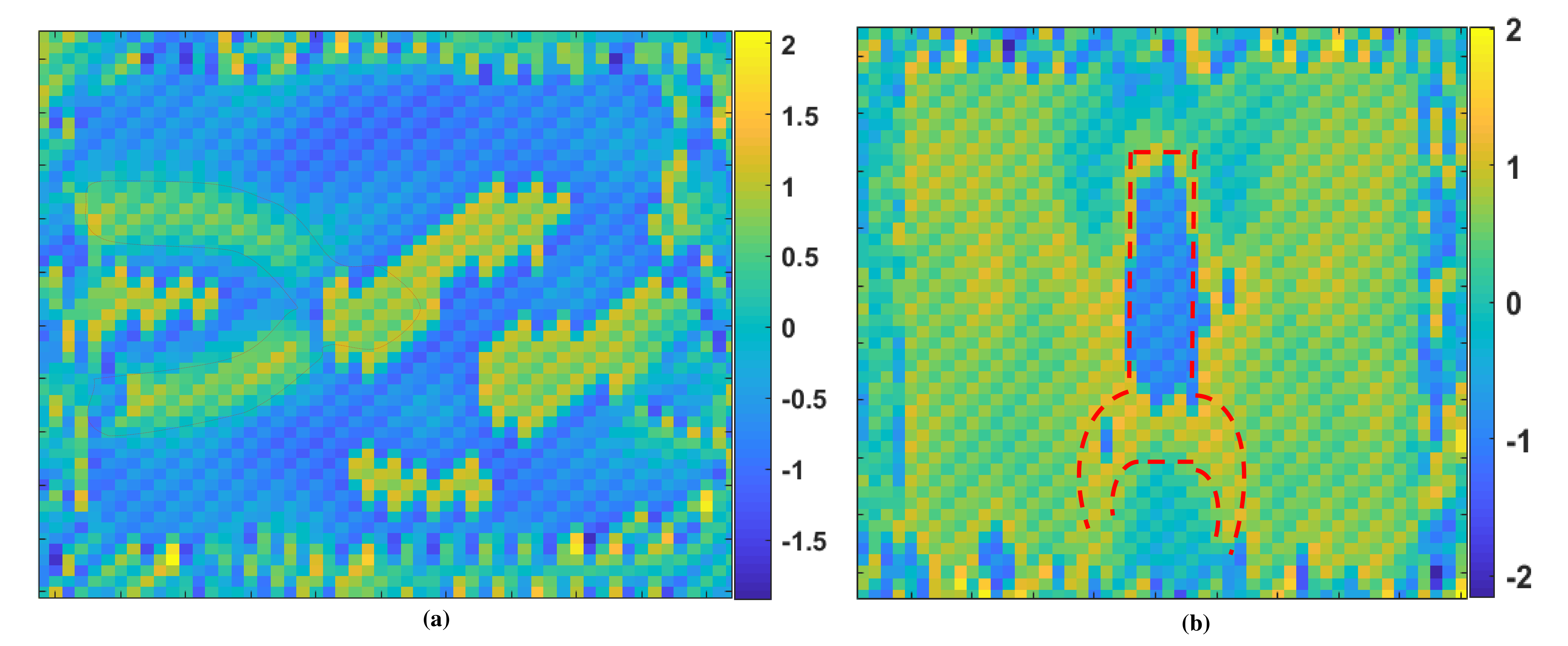}
\vspace{-0.3cm}
\caption{Reconstructed wrapped phase images of objects (a) wire scrapper (b) Spanner.}\label{f11}
\vspace{-0.3cm}
\end{figure}

The +1 order from the frequency spectrum is  filtered out and back propagated on the object plane for reconstructing the hologram using Eq. \ref{e4}. The reconstructed amplitude image is shown in Fig. \ref{f8}(a). The resolution-enhanced image is shown in Fig. \ref{f8}(b).
The reconstructed wrapped phase image is shown in Fig. \ref{f8}(c). It gives an admirable resemblance with the original object. The figures are acquainted with red marking lines to show the size and shape of the actual object. 
To investigate the ability of the Vivaldi antennae as both transmitter and receiver in the proposed indirect holographic set-up, the experiments are repeated with different metallic objects like wire scrapper and spanner. The experiments have been carried out with the same experimental parameters. The reconstructed amplitude and phase images for aforesaid objects are shown in Figs. \ref{f9}, \ref{f10} and \ref{f11}. 
In Fig. \ref{f9}(b), the amplitude image shows that the structure of wire scrapper has been reconstructed correctly. Although, the shape is in good agreement with the original object's shape, but the upper portion of the wire scrapper has not been resolved precisely because of the presence of multiple scattering sources intact closely with different heights and thickness levels. 

In Fig. \ref{f10}(a) the upper portion of the spanner is not covered under the scanning space, and it is noticed in the reconstructed image in Fig. \ref{f10}(b) that the spanner outside the scanning area has not been detected.
Thus, it can be inferred that due to the high directivity of Vivaldi antenna, the scattering field from the object only has been received and reconstructed. The obtained reconstructed images have been shown with the reconstructed images for visual comparison as in Figs. \ref{f9}(c) and \ref{f10}(c).
The phase for the objects are also reconstructed and the resultant phase images of wire scrapper and spanner are shown in Figs. \ref{f11}(a) and (b). The reconstructed phase in the following images is in much similitude with the shape of the original objects. The testing parameters of resolution enhanced holographic images are calculated utilizing Eq. \ref{e7} and \ref{e8} and are shown in Table 2.
%

\begin{figure}[htb!]
\centering
\includegraphics[width=0.8\linewidth]{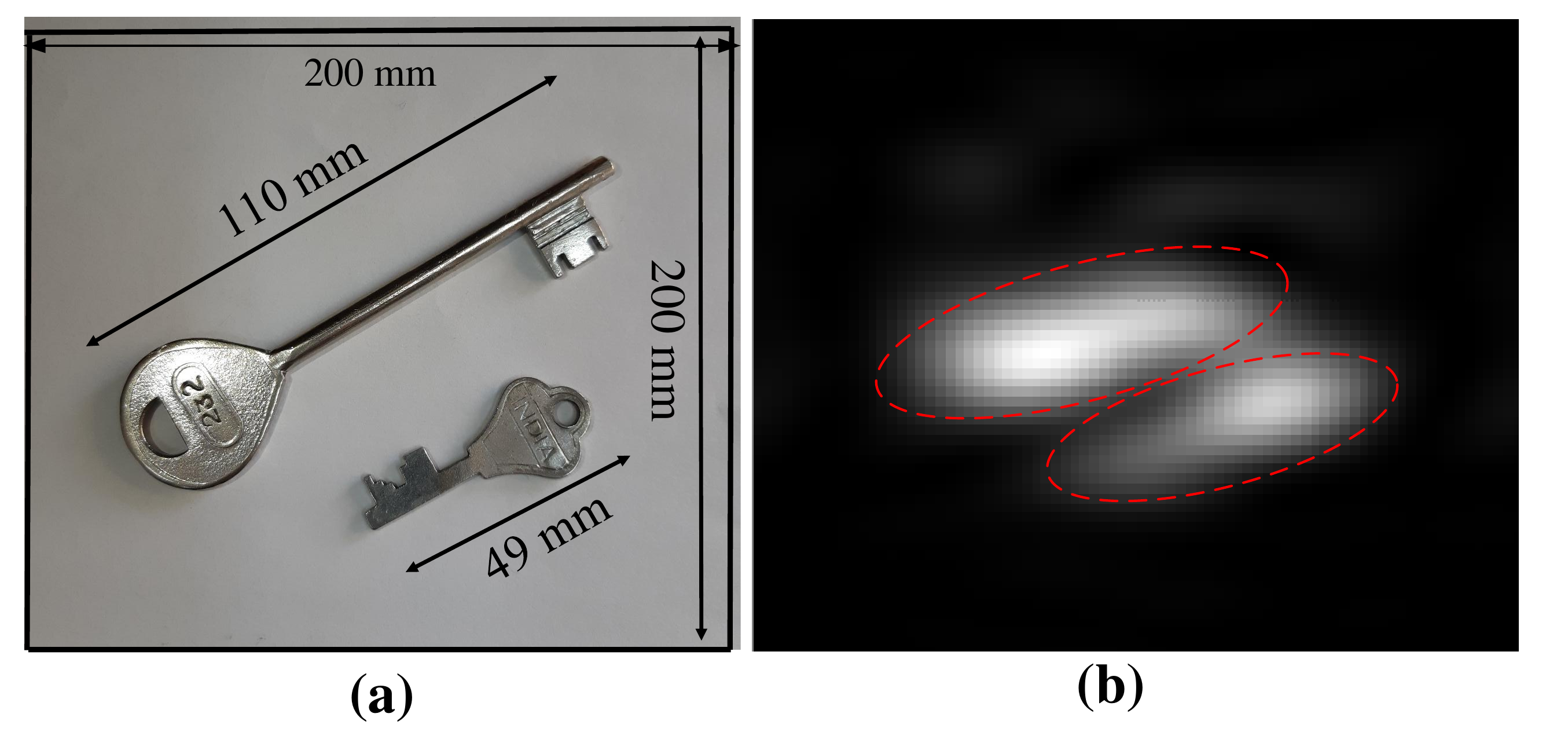}
\vspace{-0.3cm}
\caption{(a) Object image (b) Reconstructed Amplitude image.}\label{f12}
\vspace{-0.3cm}
\end{figure}

Moreover, to investigate the ability of the experimental setup for separating two close objects, two steel keys of different sizes, each of 110mm and 49mm respectively with a separation of 30mm are used as object. The reconstructed results are shown in Fig. \ref{f12}(b). The reconstructed image shows the presence of two keys with a proper separation. These reconstructed images show  that intensity difference as expected but the exact shape of the keys is not reconstructed because the dimension of the lower part of each key is of the order single pixel i.e. (5mm or less).

Post experimentation, it has been noted that however the performance of the Vivaldi antenna is good but as receiver it depends on some factors which needed to be taken care of. 
The transmitter and receiver provide the best result when both of them are having common field of view. In such geometrical conditions the perturbed scattering signals can be avoided. However, in case of microwaves it is bit difficult to do the holographic experimental alignment as microwaves are not visible to human eye. So, one should know the properties of the  antenna and then can find the best location by geometry and repeated experimental realisation. Further, the experimental arrangement must be aligned to satisfy the condition of interference. Failing to which  no fruitful reconstruction of holograms can be done. 

\section{Conclusion}
The implementation of Vivaldi antennae with a maximum gain at ${45^o}$, as transmitter and receiver in an indirect microwave holography is proposed. The design concept is novel, economic, compact, simple and promising, perceived with a combination of reconstruction and super resolution method. This work has given details of various experiments in the proof of concept while utilizing the small size directive antennae for holographic system to reduce the size of the overall set-up. The qualitative results show that the usage of Vivaldi antennae as transmitter and receiver has a significant potential in imaging applications of indirect microwave holography. Here, we have measured and demonstrated the results for metallic objects. The proposed microwave holographic set-up could be utilized and modified in terms of designing a specific antenna for various applications e.g. metal detection, biomedical imaging for cancer detection etc. This technique can be further ameliorated with more precise antenna structures and experimental design for better holographic recording and reconstruction.

\section*{Acknowledgement}
\textbf{Funding} - This research did not receive any specific grant from funding agencies in the public, commercial, or not-for-profit sectors.

\textbf{Disclosure of conflict of interest} - The authors have no relevant conflicts of interest to disclose.

\begin{table}[htb!]
  \centering
  \caption{Antenna Design Parameters for fig \ref{f1}(a) and (b)}
    \begin{tabularx}{0.6\linewidth}{>{\centering\arraybackslash}X>{\centering\arraybackslash}X}
    \toprule
    Antenna Design Parameters & Values (in mm) \\
    \midrule
    a     & 3.86 \\
    \midrule
    b     & 4.55 \\
    \midrule
    c     & 4.8 \\
    \midrule
    d     & 5 \\
    \midrule
    e     & 19.93 \\
    \midrule
    f     & 3.51 \\
    \midrule
    g     & 20.01 \\
    \midrule
    h     & 0.8 \\
    \midrule
    i     & 13.54 \\
    \midrule
    J     & 11.6 \\
    \midrule
    K     & 10.36 \\
    \midrule
    M     & 5.3 \\
    \midrule
    N     & 6.9 \\
    \midrule
    O     & 5.2 \\
    \midrule
    P     & 6 \\
    \bottomrule
    \end{tabularx}%
  \label{tab:addlabel}%
\end{table}%

\begin{table}[htbp]
  \centering
  \caption{Quantitative evaluation measures for resolution enhanced holograms}
    \begin{tabularx}{\linewidth}{XXXX}
    \toprule
    Object & \multicolumn{1}{p{8.785em}}{SNR before resolution enhancement} & \multicolumn{1}{p{8.785em}}{SNR after resolution enhancement} & SSIM (resolution enhanced image) \\
    \midrule
    `X' copper sheet & 3.8834 & 8.2544 & 0.9951 \\
    \midrule
    Wire Scrapper & 4.0587 & 9.9495 & 0.992 \\
    \midrule
    Spanner & 5.5695 & 11.248 & 0.9909 \\
    \bottomrule
    \end{tabularx}%
  \label{tab:addlabel}%
\end{table}%
\listoffigures

\end{document}